\documentclass[aps,pra,superscriptaddress,showpacs,notitlepage,twocolumn]{revtex4-1}

\usepackage{amsmath}
\usepackage{graphicx}
\usepackage{color}
\usepackage{xcolor}
\usepackage{braket}
\usepackage{verbatim}

\usepackage{amssymb}
\usepackage{amsmath}

\usepackage{soul}

\usepackage{subfigure} 
\usepackage{lipsum}%
\usepackage{hyperref}
\hypersetup{colorlinks=true}

\usepackage{graphicx}
\usepackage{dcolumn}
\usepackage{bm}

\usepackage{color}
\usepackage{braket}

\newcommand{\DFI}{Departamento de F\'isica, Facultad de Ciencias F\'isicas y Matem\'aticas, Universidad de Chile, Santiago 8370448, Chile}

\newcommand{\MIRO}{Millenium Institute for Research in Optics–MIRO, Universidad de Chile, Chile}

\begin{document}

\title{Non-symmetric evanescent coupling in photonics}

\author{Rodrigo A. Vicencio}
\affiliation{\DFI}
\affiliation{\MIRO}

\author{Diego Rom\'an-Cort\'es}
\affiliation{\DFI}
\affiliation{\MIRO}

\author{Mart\'in Rubio-Sald\'ias}
\affiliation{\DFI}
\affiliation{\MIRO}

\author{Paloma Vildoso}
\affiliation{\DFI}
\affiliation{\MIRO}

\author{Luis E. F. Foa Torres}
\affiliation{\DFI}

\begin{abstract}

Asymmetrical interactions are ubiquitous in nature, and in Physics its study becomes fundamental. Despite the prevalence of evanescent coupling in physics, little attention has been paid to wavefunction profiles, with symmetrical reciprocal interactions often assumed for practical reasons. This work challenges that assumption by analytically demonstrating the origin of non-symmetrical coupling in a photonic platform, focusing on the behavior of evanescent tails from adjacent waveguides. We experimentally validate an asymmetrical dynamics by studying detuned photonic directional couplers, fabricated via femtosecond laser writing, and corroborate our findings through continuous numerical simulations.

\end{abstract}


\maketitle

\section{Introduction} 

A fundamental concept in lattice science is the study of localized wavefunctions bounded within well-defined spatial regions, such as waveguides or atoms. The tight-binding approximation~\cite{kittel_introduction_2005} describes interactions between tightly bound orbitals at a given site that weakly interact with surrounding wavefunctions. This approach aims at examining the effects of weak coupling with neighboring sites. The approximation holds for well-separated sites where wavefunctions do not significantly overlap, as opposed to a strongly interacting regime~\cite{kaxiras_atomic_2003,SPmolecules}. In Optics, analogous equations arise from coupled mode theory (CMT)~\cite{yariv_optical_2003}, leading to key observations in transport and localization phenomena~\cite{morandotti_experimental_1999,schwartz_transport_2007,ruter_observation_2010,vicencio_observation_2015,rechtsman_photonic_2013}. Here, the inter-waveguide distance must exceed the waveguide widths to ensure minimal wavefunction distortion, allowing interactions between adjacent guided modes solely through evanescent fields. For more than two decades, research in photonic lattice phenomena~\cite{lederer_discrete_2008,ozawa_topological_2019,vicencio_poblete_photonic_2021} has predominantly assumed reciprocal coupling between waveguides. This assumption implies that the coupling constant between two waveguides is identical in both directions. Attempts to introduce non-reciprocity have mainly focused on simulating effective magnetic fields~\cite{rechtsman_photonic_2013,SebaCaging}, which add a phase to the hopping terms but do not directly alter the wavefunction itself.

A non-reciprocal directional coupler (NRDC) was originally formulated long time ago in the context of resonant circuits~\cite{berger_nonreciprocal_1965}. Then, in the 70's optical non-reciprocal devices started to be explored for integrated optics and proposals for mode converters, isolators or phase shifters were described~\cite{yamamoto_circuit_1974}. Then, the validity of CMT was a main focus of research~\cite{HS85,Haus87,marcatili1986,Huang:94}, looking for a better correspondence of the discrete model with exact numerical simulations. Some corrections to the simplest version of CMT were mandatory, with non-symmetrical coupling constants substantially different for non-identical waveguides. At the end of the 90's, magneto-optical waveguides were proposed for the construction of non-reciprocal Mach-Zehnder interferometers~\cite{bahlmann_nonreciprocal_1999}, and more recently on-chip non-reciprocal optical resonators have been implemented in silicon~\cite{bi_-chip_2011}. Time modulation has been also suggested to implement non-reciprocity in compact integrated Photonics, for example, using ring-like resonators or optomechanical cavities~\cite{sounas_non-reciprocal_2017}.

In this work, we demonstrate non-symmetrical coupling interactions in photonic detuned dimers. We develop a standard formula for coupling coefficients considering step-like index waveguides, revealing the fundamental effect of evanescent profiles from neighboring waveguides. Our analysis shows that, opposite to the standard assumption, asymmetry is indeed an inherent property of photonic coupled systems, and it originates from the difference in refractive index profiles of two distant waveguides. Using the femtosecond (fs) laser writing technique~\cite{szameit_discrete_2005}, we fabricate photonic detuned couplers and experimentally demonstrate a non-symmetrical dynamics. We corroborate all our observations through numerical simulations, without imposing any additional constraint but waveguide refractive index contrasts.

\begin{figure}[t!]
\centering
\includegraphics[width=1.\columnwidth]{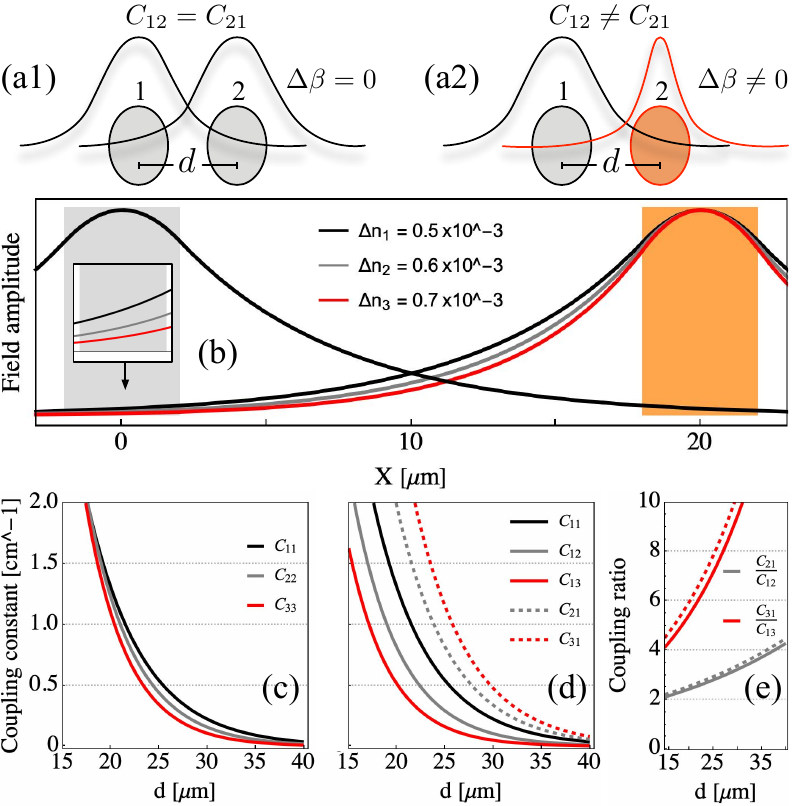}
   \caption{(a1) Homogeneous and (a2) inhomogeneous photonic dimers. (b) Dimer configuration formed by step-like index waveguides. Fixed field amplitude at the left waveguide ($\Delta n_1$) and three different mode profiles at the right waveguide ($\Delta n_1$, $\Delta n_2$ and $\Delta n_3$). Coupling constant versus distance for (c) homogeneous and (d) inhomogeneous dimers. (e) Coupling ratio versus distance. Color code is indicated in (b).}
\label{fig1}
\end{figure}
\section{Non-symmetrical photonic coupler model} 

Let us start by studying a photonic coupler, as the one sketched in Fig.~\ref{fig1}(a), consisting on two waveguides separated a distance $d$. Coupled mode theory~\cite{Okamoto} generates different integrals which account for effective field interactions originated from the assumption that the total field is written as a linear combination of the modes at waveguides $1$ and $2$: $E_T=E_1+E_2$. By assuming one-dimensional step-like index waveguides, we find different TE$_{00}$ modes, as the ones sketched in Fig.~\ref{fig1}(b), for the indicated refractive index contrasts $\Delta n_i$. In this case, mode profiles at the waveguide region (core) are proportional to ``$\cos(\alpha_i x)$'', while outside (cladding) they decay as ``$\exp(-\gamma_i |x|)$''. Fields are better confined (larger $\alpha,\gamma$) for a growing $\Delta n$, and the neighbor field will be larger at the stronger waveguide [see Fig.~\ref{fig1}(b)-inset]. Therefore, we naturally expect to observe an asymmetric coupling interaction $C_{ij}\neq C_{ji}$ for inhomogeneous couplers ($\Delta n_i\neq \Delta n_j$), with $C_{ij}$ defined as~\cite{Okamoto}
\[
C_{ij}=\omega \epsilon_0\iint (n^2-n_j^2) \vec{E}_j \vec{E}_i^* dx=\omega \epsilon_0 \Delta n_i^2 \int_{i}\vec{E}_j \vec{E}_i^* dx.
\]
$C_{ij}$ corresponds to a superposition integral at waveguide $i$, which measures the presence of mode $E_i$ and the tail coming from a neighbour waveguide mode $E_j$ ($i$ and $j$ are related to $\Delta n_i$ and $\Delta n_j$, respectively). 

First of all, we compute $C_{ii}$ for homogeneous couplers for different refractive index contrasts and show the compiled results in Fig.~\ref{fig1}(c). As it is expected, $C_{ii}$ decreases as $\Delta n_i$ increases. Now, we compute the coupling coefficients for inhomogeneous couplers, as the one sketched in Fig.~\ref{fig1}(a2). We immediately notice two strong effects that may define a \textit{non-symmetrical coupling interaction}. First of all, the mode asymmetry defines a larger coupling at the stronger waveguide but, also, the coupling directly depends on the respective index contrast $\Delta n_i$. Therefore, \textit{the asymmetry of coupling coefficients is not a weak effect and it can not be neglected when considering detuned waveguides}~\cite{HS85}. We compute coupling coefficients with respect to a reference waveguide $\Delta n_1$ (black lines) and show the results in Fig.~\ref{fig1}(d) [gray and red lines are related to $\Delta n_2$ and $\Delta n_3$, respectively]. We notice that the coupling ($C_{1j}$) experienced by this reference waveguide decreases for an increasing contrast of the neighbor waveguide $j$. This is due to the fact that the neighbor mode tail becomes weaker and weaker at the reference waveguide region [see Fig.~\ref{fig1}(b)-inset]. On the other hand, when we ask the opposite question, we find quite clearly that the coupling constant $C_{j1}$ increases for an increasing contrast $\Delta n_j$. 

We analytically compute a coupling constant ratio~\cite{SM} to estimate the strength of this effect and show the results in Fig.~\ref{fig1}(e). We observe a very strong asymmetry which grows quite fast with the separation distance $d$. We analytically approximate this ratio as
%
\begin{equation} 
\frac{C_{ji}}{C_{ij}}\approx \left(\frac{\Delta n_j}{\Delta n_i}\right)^2 e^{(\gamma_j-\gamma_i) d}\ ,
\label{ratio}
\end{equation}
and plot it with dashed lines in Fig.~\ref{fig1}(e). This expression states that two sites will experience a reciprocal and symmetric coupling interaction $C_{ij}=C_{ji}$ if and only if waveguides are equal or are able to balance this ratio. Coupling constants are the base of CMT for directional couplers~\cite{yariv_optical_2003}, and almost all the results in lattice research have been developed under a symmetric assumption, which for example keeps the standard Power and the Hamiltonian as conserved quantities~\cite{lederer_discrete_2008}.

However, equation (\ref{ratio}) also gives a strong and fundamental physical result. For sites having different refractive index contrasts, which consequently support spatially different wavefunctions, we immediately notice that $C_{ij}\neq C_{ji}$. Therefore, non-symmetrical coupling interactions emerge naturally in a photonic system. Specifically, if $\Delta n_1<\Delta n_2$, the wavefunction at site $1$ will be wider than the one at site $2$ with $\gamma_1<\gamma_2$ and, consequently $C_{12}<C_{21}$. This is a counter intuitive result because strongly bounded wavefunctions (having a larger $\gamma$) will interact with a larger coupling coefficient. Our result is rather general and applies to any physical system experiencing an evanescent coupling interaction beyond Optics and Photonics.

Considering the previous analysis, a non-symmetrical photonic coupler can be described by~\cite{Okamoto,SM}
\begin{eqnarray}
   -i\frac{du_1}{dz} =\bar{\chi}_{21} u_1+\bar{C}_{12}u_2 e^{-i\Delta\beta z}\nonumber\ ,\\
   -i\frac{du_2}{dz} =\bar{\chi}_{12} u_2+\bar{C}_{21}u_1 e^{i\Delta\beta z}\ .\ \
   \label{dimer}
\end{eqnarray}
%
%
Here $u_i$ describes the electric field mode amplitude at the $i$-th site, $z$ the propagation coordinate, $\Delta\beta\equiv\beta_2-\beta_1$ defines the propagation constants detuning, $$\bar{\chi}_{ij}\equiv\frac{\chi_{ij}-C_{ij}N_{ij}}{1-|N_{ij}|^2}$$ and $$\bar{C}_{ij}\equiv \frac{C_{ij}-\chi_{ij} N_{ij}}{1-|N_{ij}|^2}\ .$$ Coefficient $\bar{\chi}_{ij}$ is also non-symmetric and depends on the integral $\chi_{ij}$, defined as $$\chi_{ij}=\omega \epsilon_0 \Delta n_i^2\iint_{i} |\vec{E}_j|^2 dx dy\ ,$$ that describes the presence of the field $j$ at waveguide $i$ (see more details in~\cite{SM}). $\bar{C}_{12}$ and $\bar{C}_{21}$ correspond to modified coupling constants or, simply, coupling terms. Integral $N_{12}=N_{21}$ expresses a non-orthogonal contribution for the superposition of modes coming from different waveguides~\cite{Haus87,NONORT22}, and it can be written as $$N_{ij}=\frac{1}{\omega\mu_0}(\beta_i+\beta_j)\int_{-\infty}^{\infty}E_{j}E_{i}^* dx\ ,$$ for the step-like index TE$_{00}$ case. The normalization condition $N_{ii}=1$ is imposed in this derivation, and we find that $N_{12}<0.3\ll 1$~\cite{SM} and, therefore, in practice $$\bar{\chi}_{ij}\approx\chi_{ij}\ \ \text{and}\ \ \bar{C}_{ij}\approx C_{ij}\ .$$
Model (\ref{dimer}) describes a completely general dimer model including effective site detuning and non-symmetrical coupling. Although model (\ref{dimer}) is non-Hermitian, the eigenvalues are real and given by $$\bar{\lambda}_{\pm}=\pm\bar{\lambda}_0=\pm\sqrt{(\Delta\beta+\bar{\chi}_{21}-\bar{\chi}_{12})^2/4+\bar{C}_{12}\bar{C}_{21}}\ .$$ A standard power definition~\cite{lederer_discrete_2008} is not a conserved quantity of model (\ref{dimer}), but the \textit{non-symmetric power} $P_r\equiv P_1+\bar{C}_{12}P_2/\bar{C}_{21}$ is, with $P_i=|u_i|^2$. The excitation of waveguide $i$ generates a power evolution $$P_{i}^i(z)=\cos^2(\lambda_0 z)+(\Delta\beta/2\lambda_0)^2\sin^2(\lambda_0 z)$$ for the excited waveguide, and $$P_{j}^i(z)=(\bar{C}_{ji}/\bar{\lambda}_0)^2\sin^2(\bar{\lambda}_0 z)$$ for the non-excited waveguide $j$ ($i\neq j$)~\cite{SM}. Using this, we obtain a coupling ratio from dynamics given by
\begin{equation} 
\frac{P_{2}^1(z)}{P_{1}^2(z)}=\left(\frac{\bar{C}_{21}}{\bar{C}_{12}}\right)^2\approx \left(\frac{C_{21}}{C_{12}}\right)^2\ .
\label{ratio2}
\end{equation}
Therefore, depending on the excited site, asymmetry on coupling terms immediately produces an asymmetric dynamics. Specifically, for $C_{21}>C_{12}$ the exchange of energy will be more efficient when exciting waveguide $1$, as we will show below. 

\begin{figure}[t!]
\centering
\includegraphics[width=1.\columnwidth]{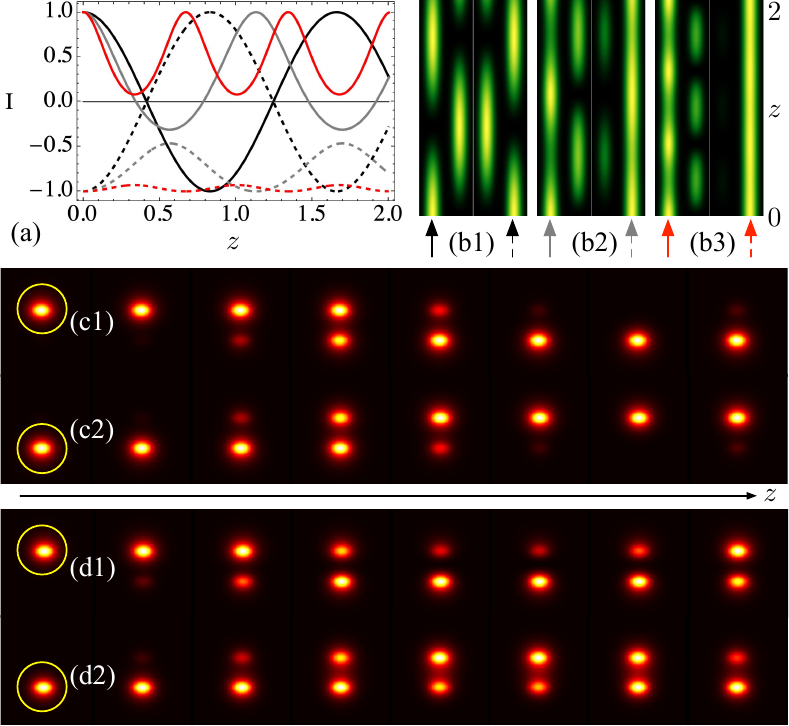}
   \caption{(a) Imbalance dynamics from model (\ref{dimer}) for the excitation of the first (full line) and second (dashed line) waveguides for $\Delta n_1-\Delta n_1$ (black), $\Delta n_1-\Delta n_2$ (gray), and $\Delta n_1-\Delta n_3$ (red) configurations. (b) Normalized power evolution, with arrows indicating the input positions and respective case relative to (a). (c) and (d) Intensity profiles versus $z$, obtained from continuous simulations, after exciting the (c1,d1) first and the (c2,d2) second waveguides. $d=18\ \mu$m.}  
\label{fig2}
\end{figure} 

In order to study the effect of this asymmetry in the dynamics, we define the \textit{Imbalance} as $$I_i=(P_1^i-P_2^i)/(P_1^i+P_2^i)\ .$$ For a reciprocal-symmetrical system the imbalance must be symmetric and the excitation of the first or the second waveguides should show mirror-symmetric curves only, independent on detuning~\cite{guzman-silva_experimental_2021}. We first explore a symmetric-reciprocal dimer, as the one shown in Fig.~\ref{fig1}(a1), with the two waveguides having a contrast $\Delta n_1$. Results are shown in Figs.~\ref{fig2}(a)-black and (b1) for $\bar{C}_{12}=\bar{C}_{21}=1.89$ cm$^{-1}$, $\bar{\chi}_{12}=\bar{\chi}_{21}=-0.24$ cm$^{-1}$ and $\Delta\beta=0$. We observe a fully symmetric energy exchange, as expected for a reciprocal directional coupler~\cite{yariv_optical_2003,Okamoto}. Then, we asymmetrize the system [see sketch in Fig.~\ref{fig1}(a2)] by taking a second waveguide with $\Delta n_2$, such that $\bar{\chi}_{21}/\bar{\chi}_{12}=2.02$, $\bar{C}_{21}/\bar{C}_{12}=2.28$, and $\Delta\beta=4.24$ cm$^{-1}$. In this case the dynamics [see Figs.~\ref{fig2}(a)-gray and (b2)] expresses a small asymmetry. We observe how the excitation of the first waveguide crosses the symmetrical point $I=0$ (i.e., a $50/50$ beam splitter) but it is not able to transfer all the power into the second waveguide. Alone, this is not a signature of non-symmetrical dynamics. But, in addition, we observe that the excitation of the second waveguide does not follow a mirror-reflected trajectory and, in fact, it is not able to even reach the symmetrical point [see Fig.~\ref{fig2}(a) dashed gray line]. Both considerations define a non-symmetrical dimer. A stronger non-symmetrical case is shown in Figs.~\ref{fig2}(a)-red and (b3) with $\bar{\chi}_{21}/\bar{\chi}_{12}=3.54$, $\bar{C}_{21}/\bar{C}_{12}=4.85$, and $\Delta\beta=8.86$ cm$^{-1}$. We observe that the excitation of the first (softer) waveguide shows a close to zero imbalance, while the excitation of the second (stronger) site shows an almost null transfer. These examples show that the observation of a Hadamard regime ($\Delta\beta=2C_{12}=2C_{21}$)~\cite{Polino:19,roadmap_2022,meany_laser_2015} is indeed not possible on a non-symmetrical coupler, due to the inherent asymmetry of this system. 

\begin{figure*}[t!]
\centering
\includegraphics[width=2.05\columnwidth]{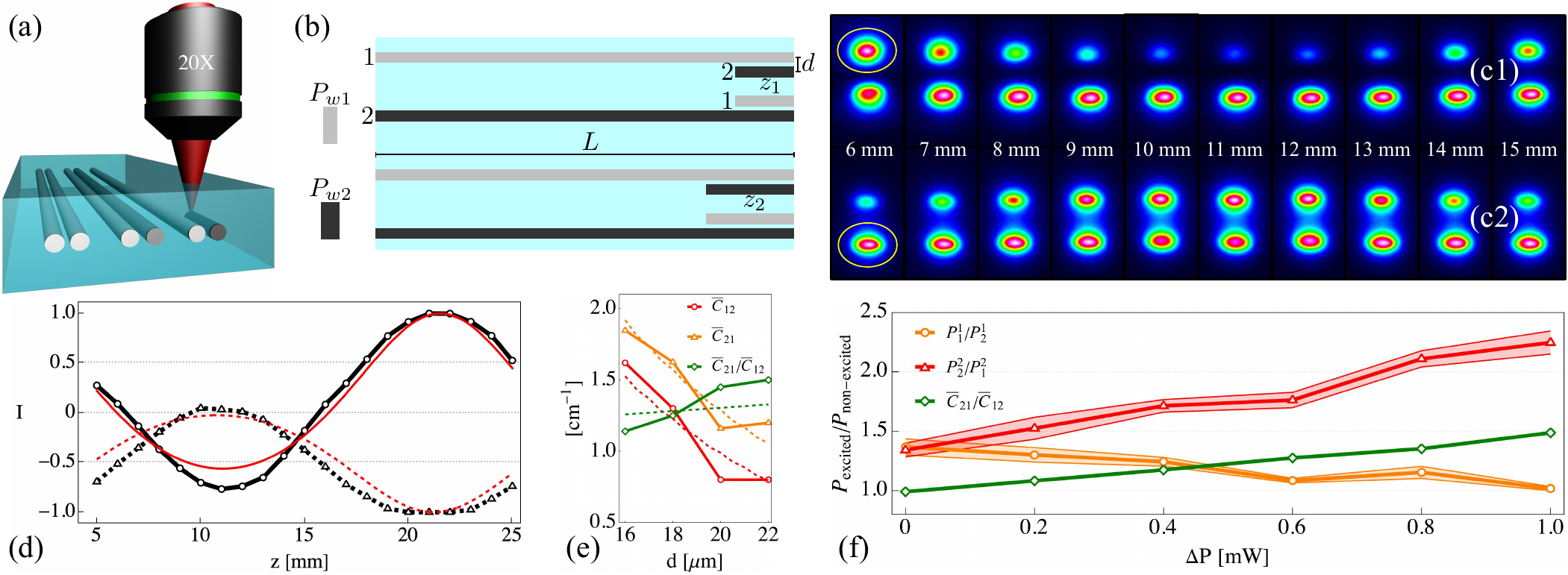}
   \caption{(a) Sketch of the femtosecond laser writing technique. (b) z-scan method for detuned couplers, fabricated with powers $P_1$ (gray stripes) and $P_2$ (black stripes), for a propagation distance $z_i$. Output intensity images versus $z_i$ for the excitation of the (c1) top and (c2) bottom waveguides, for $d=18\ \mu$m. (d) Experimental extracted imbalance versus $z_i\in\{5,25\}$ mm for (c1) black-full and (c2) black-dashed. Red lines in (d) correspond to a model (\ref{dimer}) prediction. (e) Experimentally obtained couplings $\bar{C}_{12}$, $\bar{C}_{21}$ and ratio $\bar{C}_{21}/\bar{C}_{12}$ versus distance $d$, for $\Delta P=1$ mW. Dashed lines in (e) show results from continuous simulations. (f) Average power ratio $P_i^i/P_j^i$ and its standard deviation versus $\Delta P$. Orange and red data correspond to the excitation of waveguide $1$ and $2$, respectively. Green data shows the extracted non-symmetric coupling ratio according to (\ref{ratio2}).}  
\label{fig3}
\twocolumngrid
\end{figure*}

We check model (\ref{dimer}) by means of a continuous approach~\cite{SM} and an \textit{Eigenmode Expansion Method}. As the dynamics is governed by an initial projection on linear modes, the computation of them allows us to obtain the evolution of the electric field along the propagation coordinate $z$. Figs.~\ref{fig2}(c) and (d) show examples of the obtained dynamics for a symmetric ($\Delta n_a-\Delta n_a$) and a non-symmetric ($\Delta n_a-\Delta n_b$) configuration, respectively, for $\Delta n_a=0.4\times 10^{-3}$ and $\Delta n_b=0.44\times 10^{-3}$. We observe that an homogeneous and reciprocal coupler follows a standard symmetric oscillation, with an opposite imbalance~\cite{SM}. On the other hand, an inhomogeneous coupler shows an asymmetric dynamics where the excitation of the softer waveguide [Fig.~\ref{fig2}(d1)] produces a larger energy transfer compared to the excitation of the stronger site [Fig.~\ref{fig2}(d2)]. Simulations in Figs.~\ref{fig2}(c) and (d) were performed without any constraint but a definition of waveguides having a given refractive index contrast.

\section{Experimental observation of non-symmetrical dynamics} 

Now, we provide experimental evidence for non-symmetrical coupling interactions. We fabricate several detuned photonic dimers using the fs laser writing technique~\cite{szameit_discrete_2005} on a borosilicate glass wafer, as sketched in Fig.~\ref{fig3}(a). A way to achieve detuned waveguides is by directly varying the fabrication parameters; i.e., laser power and/or writing velocity. The larger the power and the slower the velocity the stronger the refractive index change~\cite{szameit_discrete_2005,guzman-silva_experimental_2021}. We set the velocity to $v=2.5$ mm/s and the writing power $P_{w1}$ for waveguide $1$. Non-symmetric couplers are fabricated by applying a larger writing power $P_{w2}$ for waveguide $2$. In all the experiments, we collect intensity images with a beam profiler at the output facet of a $L=30$ mm long wafer. The dynamics along the $z$ coordinate can be studied by conducting a $z$-scan experiment~\cite{guzman-silva_experimental_2021}: dimers are fabricated in pairs with a full waveguide of length $L$ plus a shorter waveguide of length $z_i$, separated by an inter-site distance $d$, as sketched in Fig.~\ref{fig3}(b), such that we can excite the first and the second waveguide of an equivalent coupler.

The $z$-scan experiment runs in the interval $z_i\in\{5,25\}$ mm, every $1$ mm, and show a shorter set of images in Fig.~\ref{fig3}(c). In this case, bottom waveguides were fabricated such that $\Delta P\equiv P_{w2}-P_{w1}=5$ mW, for $d=18\ \mu$m. Fig.~\ref{fig3}(c1) shows an out of phase dynamical evolution after exciting waveguide $1$, where we observe an almost full energy transfer at $z\approx 11$ mm [$I\sim -0.75$ in Fig.~\ref{fig3}(d)]. On the other hand, $\sim 50\%$ of the energy is transferred by exciting waveguide $2$ as shown in Fig.~\ref{fig3}(c2), with $I\sim 0$ in Fig.~\ref{fig3}(d). The images shown in Fig.~\ref{fig3}(c), and the compiled Imbalance data of Fig.~\ref{fig3}(d)-black, give a clear experimental proof of non-symmetrical dynamics, with asymmetric trajectories after initializing the system at opposite waveguides. We perform a calibration of parameters by fitting analytical dimer solutions~\cite{SM,HS85} from model (\ref{dimer}) with the data shown in Fig.~\ref{fig3}(c). Results are shown with red lines in Fig.~\ref{fig3}(d), for $\Delta\beta+\bar{\chi}_{21}-\bar{\chi}_{12}\approx 1.8$ cm$^{-1}$, $\bar{C}_{12}\approx 0.85$ cm$^{-1}$, and $\bar{C}_{21}\approx 1.70$ cm$^{-1}$. This example shows an indeed strong asymmetry with a coupling ratio $\bar{C}_{21}/\bar{C}_{12}\approx 2.0$. 

We implement several experiments to confirm this very fundamental observation. For example, we set a smaller power detuning $\Delta P=1$ mW and fabricate several couplers for different inter-site distances $d$, including a $z$-scan experiment to extract the coupling parameters. Fig.~\ref{fig3}(e) shows our compiled results for coupling constants, where we observe a standard decreasing dependence~\cite{szameit_discrete_2005,guzman-silva_experimental_2021}. We also notice a clear gap in between $\bar{C}_{12}$ and $\bar{C}_{21}$, confirming the non-symmetrical nature of the detuned photonic couplers. We compute the ratio $\bar{C}_{21}/\bar{C}_{12}$ (see green full line) and observe an increasing tendency while increasing the inter-site distance, in agreement with prediction (\ref{ratio}). In addition, we confirm these experimental findings by continuous numerical simulations [see dashed lines in Fig.~\ref{fig3}(e)], assuming detuned couplers with a small refractive index difference $\Delta n\equiv n_2-n_1=0.2\times 10^{-4}$.


Finally, we confirmed statistically the observed non-symmetrical dynamics by repeating the dimer experiments several times, for a fixed set of parameters: $P_{w1}$, $d=20\ \mu$m, and $z_1=6$ mm. We fabricate a set couplers pairs, where the long and short waveguides are simply exchanged [see Fig.~\ref{fig3}(b)]. We fabricate the clone systems 10 times and obtain a rough statistical confirmation for different detuned waveguides with an increasing power detuning $\Delta P$. Our compiled results are shown in Fig.~\ref{fig3}(f). As non-symmetrical interactions are dynamically evidenced by an asymmetric evolution, in this experiment we measured the output power ratio among the excited and the non-excited waveguides: $P_i^i/P_j^i$. Fig.~\ref{fig3}(f) shows our averaged results where we clearly observe a non-symmetrical dynamics, with a very small standard deviation. Only for equal waveguides ($\Delta P=0$), we observe a convergent dynamics with an equal power ratio, as expected for homogeneous coupler systems. However, once $\Delta P$ increases, we start observing a divergent dynamics: the excitation of the low power waveguide ($P_{w1}$) produces a larger energy transfer (smaller ratio), while the excitation of the higher power waveguide ($P_{w2}$) tends to keep the energy trapped at the input site (larger ratio). From (\ref{ratio2}) we compute the ratio in between both experiments and obtain a coupling ratio $\bar{C}_{21}/\bar{C}_{12}$ [see green data in Fig.~\ref{fig3}(f)], which increases linearly with the writing power $\Delta P$ (detuning).

\section{Conclusions}

In this work, we have developed the concept of non-symmetrical coupling emerging from detuned photonic waveguides. We analytically, numerically and experimentally demonstrated that detuned mode profiles lead to non-symmetric coupling constants. While evanescent coupling is a natural phenomenon in Optics, we believe this concept could be extended to other physical contexts with similar descriptions based on superposition integrals. All experimental observations were corroborated by extensive numerical simulations of a paraxial wave equation, using only realistic optical parameters without additional constraints. Our findings demonstrate that standard table-top experiments can now be used to study more complex phenomena, potentially opening new research directions and having a broad impact in physics, and science in general. Breaking the symmetry is fundamental yet challenging to achieve in compact integrated devices~\cite{sounas_non-reciprocal_2017} and our proposal offers a clear and simple solution to this fundamental challenge in applied Photonics~\cite{externalcontrolPRL21,OBrien08,meany_laser_2015,Polino:19,roadmap_2022} with, in principle, a small footprint and scalability. Our experimental proposal could be extended to non-Hermitian systems~\cite{martinez_alvarez_non-hermitian_2018,foa_torres_perspective_2020,bergholtz_exceptional_2021,yao_edge_2018,lin_topological_2023,okuma_non-hermitian_2023}, where different exotic transport properties emerge~\cite{added01,added02,added03}.

\section*{Acknowledgements.} This research was supported in part by Millennium Science Initiative Program ICN17$\_$012, ANID FONDECYT Grants 1231313 and 1211038. L.E.F.F.T. acknowledges support from The Abdus Salam International Center for Theoretical Physics and the Simons Foundation, and by the EU Horizon 2020 research and innovation program under the Marie-Sklodowska-Curie Grant Agreement No. 873028 (HYDROTRONICS Project).


%

\end{document}